\UseRawInputEncoding
\documentclass[preprint,onecolumn,nofootinbib]{revtex4}
\pdfoutput=1
\usepackage[colorlinks=true,linkcolor=blue,urlcolor=blue,filecolor=black,citecolor=red,pdfstartview=FitV,pdftitle={},pdfsubject={},pdfkeywords={},pdfpagemode=None,bookmarksopen=true]{hyperref}
\usepackage{graphicx}
\usepackage{amsmath}
\usepackage{amsfonts}
\usepackage{amssymb,ulem}
\usepackage{color,xcolor}%
\usepackage{CJK}
\usepackage{subfigure}

\usepackage{dcolumn}
\usepackage{MnSymbol,wasysym}
\usepackage{braket}

\usepackage{eurosym}
\usepackage{calrsfs}

\begin{document}
\title{Quantum fluctuation on the worldsheet of probe string in BTZ black hole }

\author{Yu-Ting Zhou}
\email{yu-tingzhou@nuaa.edu.cn}
\affiliation{College of Physics, Nanjing University of Aeronautics and Astronautics, Nanjing 210016, China}
\affiliation{ Key Laboratory of Aerospace Information Materials and Physics (NUAA), MIIT, Nanjing 211106, China}

\author{Xiao-Mei Kuang}
\email{xmeikuang@yzu.edu.cn }
\affiliation{ Center for Gravitation and Cosmology, College of Physical Science and Technology, Yangzhou University, Yangzhou 225009, China.}

\begin{abstract}
In this paper, we investigate the second-order normal quantum fluctuation on the worldsheet of a probe string in the Ba\~nados-Teitelboim-Zanelli (BTZ) black hole. These fluctuations is treated as the projection of Hawking radiation on the worldsheet and indeed modify the action growth of the string. Then in the string field theory/boundary conformal field theory framework, via the boundary vertex operator we study the correlation function of the Schr$\ddot{o}$dinger functional of excited fields on the worldsheet and further extract the field's formula. Our study could shed light on the potential connection between complexity growth and correlation function.
\end{abstract}

\maketitle
\tableofcontents

\section{Introduction}
Black holes have become a natural platform for us to understand spacetime and gravity at a semi-classical level after Stephen Hawking proposed the exciting theory of black hole radiation. Then, drawing inspiration from black hole thermodynamics, physicists proposed the holographic nature of gravity 
 \cite{Susskind:1994vu,tHooft:1999rgb}, which then gives rise to the gauge/gravity duality \cite{Maldacena:1997re,Gubser:1998bc,Witten:1998qj}. The extensive applications of this duality provide  novel perspectives for exploring gravity and strongly coupled system. 

Recently, quantum information theory has begun to transcend its traditional framework, providing more and more insights in the field of quantum gravity such as computational complexity \cite{Osborne_2012,TCS-066,Dvali000,PhysRevA.94.040302,PhysRevD.96.126001,Watrous:2008,Bao:2018ira}. The complexity essentially measures the difficulty of changing a quantum state into another state, however, it is still not clear to define the initial states and target
states when one applies complexity into  quantum field theory (QFT). Though considerable attempts have been made in this field \cite{Vanchurin:2016met,Chapman:2017rqy,Molina-Vilaplana:2018sfn,Bhattacharyya:2018wym,Nielsen:2005mkt,doi:10.1126/science.1121541,Jefferson:2017sdb,Yang:2018nda,Bhattacharyya:2018bbv,Bhattacharyya:2019kvj,Camargo:2022wkd,Adhikari:2022whf,Adhikari:2021pvv,Adhikari:2022oxr}, a widely acceptable  understanding of the complexity still poses an unresolved query. Thanks to holography, two elegant descriptions of complexity have been proposed from gravity side. One  is the ``complexity=volume (CV)'' conjecture, where V represents the volume of the Einstein-Rosen (ER) bridge linking the two sides of the AdS black hole's boundary. The other is the ``complexity=action (CA)'' conjecture, where A denotes the classical action of a space-time region enclosed by the bulk Cauchy slice anchored at the boundaries, also known as the ``Wheeler-Dewitt (WdW)" patch \cite{Chapman:2016hwi,Brown:2015bva,Brown:2015lvg}.

Based on the CA conjecture, there are many studies on the stationary systems, see for examples \cite {Pan:2016ecg,Guo:2017rul,Momeni:2016ekm,Tao:2017fsy,Alishahiha:2017hwg,Reynolds:2017lwq,Qaemmaqami:2017lzs,Sebastiani:2017rxr,Couch:2017yil,Swingle:2017zcd,Cano:2018aqi,Chapman:2018dem,Chapman:2018lsv,Auzzi:2018pbc,Yaraie:2018hwz,Alishahiha:2018tep,An:2018xhv,Cai:2016xho,Ghodsi:2020qqb,Frassino:2019fgr} and references therein. Similar efforts are evident in dynamic systems, {such as the investigation of complexity growth with probe branes \cite{Abad:2017cgl,Santos:2022lxj} and the exploration of non-local operator effects in the BTZ black hole} \cite{Ageev:2014nva,Zhou:2021vsm,Zhou:2023nza,Nagasaki:2018csh,Nagasaki:2019icm, Bravo-Gaete:2020lzs, Santos:2020xox, Nagasaki:2021ldz, Nagasaki:2022lll}. 
Besides in Andi-de Sitter (AdS) spacetime, complexity in de Sitter (dS) spacetime 
 was recently studied, and it was found that in this case  the holographic complexity exhibits `hyperfast' growth \cite{Susskind:2021esx,Jorstad:2022mls}. Later, the authors of \cite{Santos:2023eqp} investigated the phase transition between the dS and AdS spacetime regimes based on the holographic entanglement entropy and the renormalization group flow, which could provide insight to understand more holographic properties, such as complexity, in different energy scales. Moreover,
holographic complexity was found  to have potential to describe 
the information emission from the rotating  BTZ black hole encoded by quantum complexity \cite{Brown:2017jil}, even at zero temperature \cite{Santos:2024zoh}.Though considerable remarkable progress has been made in holographic complexity, a well-defined reference state in holography is also open. Therefore, the study of quantum fluctuation or perturbation of complexity could circumvent the state puzzle and help to step forward to investigate the complexity.

Thus, the aim of this paper is to study the second-order normal fluctuation on a probe string whose two points end in the dual boundary of the BTZ  black hole. As a first attempt, our goal is to build a possible connection between the complexity growth and correlation
function. The purpose we consider the normal fluctuations on the probe string mainly consists of two aspects.
On one hand,
the probe string in BTZ black hole can be treated as a 2-dimensional quantum field theory in the curved space-time background, similar as the handling with holographic Brownian motion proposed in \cite{deBoer:2008gu,Atmaja:2010uu}. Then in this scenario,  various fluctuation modes could be excited on the string's worldsheet due to Hawking radiation in the black hole environment \cite{Lawrence:1993sg}. Here for convenience,  we shall fix the end points of the string onto the boundary and focus on the normal direction fluctuation of the string. This operation allows us to obtain the normal operator of those excited normal modes along the string, as we will show soon.
On the other hand, this consideration allows us to work with the string field theory/boundary conformal field theory (SFT/BCFT) correspondence, stating that every classical field in worldsheet of the probe string can be described by a BCFT of an open string attached to the boundary \cite{Fuchs:2008cc, Taylor:2003gn, Rastelli:2005mz, Taylor:2006ye, Kiermaier:2008qu,Kudrna:2012re}. 
In particular, using the AdS/BCFT scenario in Horndeski gravity \cite{Santos:2021orr}, the authors of \cite{Santos:2024zoh} found that the corrections provided in the probe string worldsheet in a rotating BTZ black hole indeed contribute to the action growth.
In this framework, we can treat the excited fields as a single object and construct the corresponding  vertex operator on the boundary. This then further helps us to calculate the correlation function of the bulk fields in the Schr$\ddot o$dinger functional representation and extract
the curved-dependent worldsheet excited field.

The remaining of this paper is organized as follows. In Sec. \ref{sec2}, we briefly review the  BTZ black hole and the action growth of probe string model. In Sec. \ref{sec3}, we consider second order normal fluctuation on the Nambu-Goto action of the probe string in the black hole and then we get the normal fluctuation operator. In Sec.\ref{sec4}, we explore the two-point correlation function of these excited modes and extract the field function on the worldsheet. We summarize our work in the last section.

\section{Action growth of probe string in BTZ black hole}\label{sec2}
We start with the 3-dimentional BTZ black hole with a negative cosmological constant of which the metric is \cite{PhysRevLett.69.1849, PhysRevD.48.1506}
\begin{equation}
ds^{2}=-(-M+\frac{r^{2}}{l^{2}})dt^{2}+\frac{1}{-M+\frac{r^{2}}{l^{2}}}dr^{2}+r^{2}d\phi^{2},
\label{BTZbh}
\end{equation}
where $-\infty \leq t \leq \infty $ and $0 \leq \phi \leq 2\pi$, $M$ is the black hole mass and $l$ is AdS radius.  Then for convenience, we shall set $l=1$ and rewrite metric \eqref {BTZbh} under Poincare coordinate as,
\begin{equation}
ds^{2}=\frac{1}{z^{2}}\left[-f(z)dt^{2}+\frac{1}{f(z)}dz^{2}+d\phi^{2}\right],\; f(z)=1-Mz^{2},
\label{BTZbhZ}
\end{equation}
where we have $z=\frac{1}{r}$, the horizon location $z_{h}=\frac{1}{\sqrt{M}}$ and the Hawking temperature 
$T=\frac{1}{2 \pi \sqrt{M}}$. 

\begin{figure}[ht!]
 \centering
  \includegraphics[width=9cm]{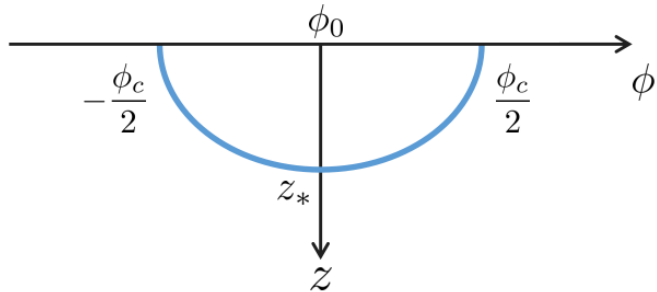}
    \caption{The geometric description of a probe string ending on the boundary of BTZ black hole background.}
 \label{configuration}
\end{figure}

We proceed to consider a probe string in this background spacetime 
and its two endpoints  are attached in  the boundary subspace. The configuration is shown in FIG. \ref{configuration} where  we have omitted the time direction.
Subsequently,  we can work in the worldsheet coordinates $\tau = t$ and $\sigma = \phi$, and further perform $X^{\mu}(t,\phi)$ as 
\begin{eqnarray}
X^{\mu}(t,\phi)= \left( \begin{matrix} &t \\& z(\phi)  \\& \phi
\end{matrix}  \right).
\label{3}
\end{eqnarray}
Then the Nambu-Goto action of this string is 
\begin{equation}
S_{NG}= T_{s} \int dt d\phi \sqrt{-\det \; [h_{ab}]} \;,
\label{4}
\end{equation}
where $T_{s}$ is the tension of string. $h_{ab}$ is the induced metric of worldsheet, which can be written in the form
\begin{equation}
h_{ab}=g_{\mu \nu}\partial_{a} X^{\mu}\partial_{b}X^{\nu},
\label{5}
\end{equation}
with $g_{\mu \nu}$ the metric defined in \eqref{BTZbhZ}.
Then directly from \eqref{4}, we can define the action growth of the probe string as 
\begin{equation}
\frac{1}{T_{s}}\frac{dS_{NG}}{dt}=\int_{\frac{-\phi_{c}}{2}} ^{\frac{\phi_{c}}{2}} d\phi \sqrt{-\det \; [h_{ab}]} \;.
\label{6}
\end{equation}
Further calculations on the above action growth has been addressed in \cite{Nagasaki:2018csh,Zhou:2023nza, Zhou:2021vsm}, in which the authors discussed the significant effects of the graviton mass and black hole mass, similar to that from CV approach \cite{Zhou:2019jlh}.  In the following studies, we shall concern the quadratic fluctuation on this action growth.

\section{Quantum fluctuation on the worldsheet of probe string}\label{sec3}
\begin{figure}[ht!]
 \centering
  \includegraphics[width=9cm]{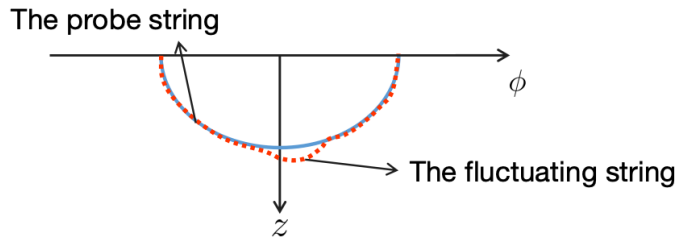}
    
    \caption{The geometric description of the fluctuation on the string.}
 \label{fig2}
\end{figure}

We shall then adopt the normal coordinate gauge, under which the fluctuation is
normal to the probe string everywhere, to  expand the action growth into the quadratic order. The reason exists in that the normal coordinate gauge could be the safest choice comparing
to other coordinate gauges, as addressed in \cite{Kinar:1999xu}. The explicit geometric description of this fluctuation is shown in FIG.\ref {fig2}, in which the dashed curve describes a fluctuated string on the original probe string.

To proceed, we expand the action growth \eqref{6} up to the quadratic term as 
\begin{equation}
\frac{1}{T_{s}}\frac{dS_{NG}}{dt}\simeq\frac{1}{T_{s}}\frac{dS_{NG}^{(0)}}{dt}+\frac{1}{T_{s}}\frac{dS_{NG}^{(2)}}{dt}+\cdots,
\label{7}
\end{equation}
where the first term in the right side denotes the initial classical action growth. It is noticed that in general we can have infinite  higher order terms  but in our following study we  consider till quadratic term.  Then we shall assume that the fluctuation field is a scalar field $\xi_{n}$ and account for the contributions from the quantum fluctuations. To this end,
we first expand the string coordinates around (\ref{3}) as
\begin{equation}
X^{\mu}(t, \phi)=X_{0}^{\mu}(t, \phi)+n_{\mu}\xi_{n}(t,\phi).
\label{8}
\end{equation}
Here we introduce the unit normal vector $n_{\mu}=(n_{z},n_{\phi})$ and consider the normal fluctuation on the coordinate. Subsequently, (\ref{8}) can be rewritten as
\begin{eqnarray}
&&z=z+n_{z}\xi_{n}(t,\phi) , \label{9} \\
&&\phi=\phi+n_{\phi}\xi_{n}(t,\phi).
\label{10}
\end{eqnarray}
Noted that there is no fluctuation on the  time direction, which means that we keep the time coordinate fixed. The normal vector should  satisfy  the condition \cite{Kinar:1999xu} 
\begin{eqnarray}
&&n_{\phi} \lambda_{\phi} g_{\phi \phi}+n_{z}\lambda_{z}g_{zz}=0 , \label{11} \\
&&n_{\phi}^{2}g_{\phi \phi}+n_{z}^{2}g_{zz}=1,
\label{12}
\end{eqnarray}
where $\lambda_{\phi}$ and $\lambda_{z}$ are the parameters satisfying the tangent vector equation of string
\begin{equation}
z^{'}=\frac{\partial z}{\partial \phi}=\frac{\lambda_{z}}{\lambda_{\phi}}.
\label{13}
\end{equation}

Now we have to fix the normal vector. The Hamiltonian of the string can be extracted from \eqref{4} as
\begin{equation}
H=\frac{M z^{2}-1}{z^{2}\sqrt{1-M z^{2}+z^{'2}}},
\label{14}
\end{equation}
which does not explicitly depend on $\phi$, so it is conserved. Then $z^{'}$ could be written as 
\begin{equation}
z^{'}=\frac{z_{*}^{2}\sqrt{Mz^{2}-1}\sqrt{Mz^{2}-1+\frac{z^{4}(1-Mz_{*}^{2})}{z_{*}^{4}}}}{z^{2}\sqrt{1-M z_{*}^{2}}},
\label{15}
\end{equation}
where $z_{*}$ is the turning point at the $\phi_0$ (see FIG.\ref{configuration}). Then combining (\ref{11})-(\ref{13}) and (\ref{15}), we can solve out $n_{z}$ and $n_{\phi}$ as
\begin{eqnarray}
&&n_{z}=\frac{z^{3} \sqrt{1-M z_{*}^{2}}}{z_{*}^{2}}, \notag \\ 
&&n_{\phi}=\frac{\sqrt{z^{6}-Mz^{6}z_{*}^{2}-z^{2}z_{*}^{4}+Mz^{4}z_{*}^{4}}}{\sqrt{Mz^{2}z_{*}^{4}-z_{*}^{4}}}.
\label{16}
\end{eqnarray}
Under the fluctuated string coordinates
\begin{eqnarray}
&&\tilde{X}^{t}=t, \notag \\
&&\tilde{X}^{z}=z+n_{z}\xi_{n}(t,\phi), \notag \\
&&\tilde{X}^{\phi}=\phi+n_{\phi}\xi_{n}(t,\phi), 
\label{17}
\end{eqnarray}
the induced metric \eqref{5} can be expanded as
\begin{eqnarray}
&&\tilde{h}_{tt}=\partial_{t}\tilde{X}^{t}\partial_{t}\tilde{X}^{t}g_{tt}+\partial_{t}\tilde{X}^{z}\partial_{t}\tilde{X}^{z}g_{zz}+\partial_{t}\tilde{X}^{\phi}\partial_{t}\tilde{X}^{\phi}g_{\phi \phi}, \notag\\
&&\;\; \; \; \; \;=g_{tt}+(n_{z}\partial_{t}\xi_{n})^{2}g_{zz}+(n_{\phi}\partial_{t}\xi_{n})^{2}g_{\phi \phi}  \notag  \\
&& \tilde{h}_{\phi \phi}=\partial_{\phi}\tilde{X}^{z}\partial_{\phi}\tilde{X}^{z}g_{zz}+\partial_{\phi}\tilde{X}^{\phi}\partial_{\phi}\tilde{X}^{\phi}g_{\phi \phi} \notag \\
&&\;\; \; \; \; \;=[z'+\partial_{\phi}(n_{z}\xi_{n})]^{2}g_{zz}+[1+\partial_{\phi}(n_{\phi}\xi_{n})]^{2}g_{\phi \phi}\notag  \\
&&\;\; \; \; \; \;=(z')^{2}g_{zz}+2z'\xi_{n}\partial_{\phi}n_{z}\cdot g_{zz}+2z'n_{z}\partial_{\phi}\xi_{n}\cdot g_{zz}+\xi_{n}^{2}(\partial_{\phi}n_{z})^{2}g_{zz} \notag \\
&&\;\; \; \; \; \;+2\xi_{n}n_{z}\partial_{\phi}n_{z}\cdot \partial_{\phi}\xi_{n}\cdot g_{zz}+n_{z}^{2}(\partial_{\phi}\xi_{n})^{2}g_{zz}+g_{\phi \phi}+2\xi_{n}\partial_{\phi}n_{\phi}\cdot g_{\phi \phi} \notag \\
&&\;\; \; \; \; \;+2n_{\phi}\partial_{\phi}\xi_{n}\cdot g_{\phi \phi} +\xi_{n}^{2}(\partial_{\phi}n_{\phi})^{2}g_{\phi \phi}+2\xi_{n}n_{\phi}\partial_{\phi}n_{\phi}\cdot \partial_{\phi}\xi_{n}\cdot g_{\phi \phi}+n_{\phi}^{2}(\partial_{\phi}\xi_{n})^{2}g_{\phi \phi}.
\label{18} 
\end{eqnarray}
After straightforward calculation, we can expand the $\det [\tilde{h}_{ab}]$ up to the second order and obtain the quadratic term
\begin{eqnarray}
\frac{1}{T_{s}}\frac{dS_{NG}^{(2)}}{dt}=\int d\phi \;\; \xi_{n}^{\dagger} \left[ (z'^{2}g_{zz}+g_{\phi \phi})\partial_{t}^{2}+\partial_{\phi}(g_{tt}\partial_{\phi})-[(\partial_{\phi}n_{z})^{2}g_{zz}g_{tt}+(\partial_{\phi}n_{\phi}^{2})g_{\phi \phi}g_{tt}]\right]\xi_{n},
\label{19}
\end{eqnarray}
where we have used the relation of the normal vector in (\ref{13}). By defining a normal fluctuation operator as 
\begin{eqnarray}
O_{n}=(z'^{2}g_{zz}+g_{\phi \phi})\partial_{t}^{2}+\partial_{\phi}(g_{tt}\partial_{\phi})-[(\partial_{\phi}n_{z})^{2}g_{zz}g_{tt}+(\partial_{\phi}n_{\phi})^{2}g_{\phi \phi}g_{tt}],
\label{20}
\end{eqnarray}
we can rewrite the  quadratic terms \eqref{19}   as 
\begin{equation}
\frac{1}{T_{s}}\frac{dS_{NG}^{(2)}}{dt}= \sum^{N}_{n} \int d\phi \xi_{n}^{\dagger}O_{n} \xi_{n},
\label{21}
\end{equation}
which was proposed in \cite{Kinar:1999xu}.

In principle, we can solve out  $\xi_{n}$ from (\ref{21}) by the equation of motion
\begin{eqnarray}
O_{n}\xi_{n}=0, 
\label{22}
\end{eqnarray}
however, it is a very difficult mission for us to go ahead. The difficulties come from two aspects : the first is  that $\xi_{n}$ here are not a single field but a collection of all possible fields, so we cannot solve it by the separation method proposed for single filed \cite{Kinar:1999xu} of which  the authors  treated it as an eigenvalue problem and separated $\phi(v,t)$ into $e^{i\omega t}\phi(v)$ to solve it.
The second is that the equation $O_{n}\xi_{n}$ is highly complex, so analytical solution is a big challenge and numerical method is called for. Therefore, we shall shift our strategy and adopt an alternative approach to find $\xi_{n}$, which will be elaborated in the next section.

\section{Further study from the perspective of String Field Theory}\label{sec4}
In this section, we shall endow the quantum fluctuation on the probe string with a potential physical process and discuss what we can construct. The probe string living in BTZ black hole environment should be inevitably affected by the Hawking radiation \cite{deBoer:2008gu,Atmaja:2010uu,Lawrence:1993sg}, besides,  combining (\ref{8}) and (\ref{21}) further inspires us to interpret  $\xi_{n}$ as the fluctuation fields caused by the Hawking radiation.
It is noted that random fluctuations on the worldsheet caused by Hawking radiation can also lead to a random motion of the endpoint of the string on the boundary, which is dubbed  holographic Brownian motion \cite{deBoer:2008gu}. But here our case is different because we fix the two endpoints of the string on the boundary and only consider the fluctuation on the worldsheet. Then one interesting question we shall ask is which kind of physical quantity in the fluctuations can correspond to complexity/action growth caused by this process. We shall do some analysis and expect to give insight in this issue.

As we aforementioned that the quantum fluctuations could be excited by any kind of field so that the sum in (\ref{21}) should collect all possible fields. 
Fortunately, inspired by the strategy in  string field theory (SFT) \cite{Taylor:2003gn,Fuchs:2008cc,Erler:2019vhl}, we  can treat this set of fluctuations as a single object, called excited string fields (ESF) $\Xi[\phi_{i},A_{\mu}, ...]$, where $\phi_{i}$ and $A_{\mu}$ are the scalar fields and  $U(1)$ gauge field, while the ellipsis denote other possible  excited fields. 
Moreover, in our setup the worldline of the string is swept out by the endpoints attached to the boundary, so all of these components collectively form boundary conformal field theory. 
Thus, we shall apply SFT/BCFT correspondence \cite{Fuchs:2008cc, Taylor:2003gn, Rastelli:2005mz, Taylor:2006ye, Kiermaier:2008qu,Kudrna:2012re} to proceed. 

We define a worldsheet's excited space , $\mathcal{H}_{BCFT}$ , which can be viewed as a tensor productor of ``matter " and ``ghost" sectors \cite{Polchinski}
\begin{equation}
\mathcal{H}_{BCFT}=\mathcal{H}^{m}_{BCFT} \otimes \mathcal{H}^{gh}_{BCFT},
\label{23}
\end{equation} 
where $m$ and $gh$ denote matter and ghost sectors, respectively. The ghost sector is a $bc$ system that is characterized by anticommuting, holomorphic, and anti-holomorphic worldsheet fields $b,\bar b, c,\bar c$. Since the ghost sector comes from the gauge fixing process and its physics is usually not well understood, so here  we do not consider the ghost sector for the fluctuations. 
For a state $\ket{\xi_{n}} \in \mathcal{H}_{BCFT}$ , there exits a corresponding boundary operator $V_{\xi_{n}}$, called vertex operator, so that we have 
\begin{equation}
\ket{\xi_{n}}=V_{\xi_{n}}\ket0 .
\label{24}
\end{equation}
Here $\ket0$ is the $SL$(2, $\mathbb R$) vacuum in the boundary without vertex operator. Then we extend the vertex operator at the point $\phi_{0}$ as $\mathcal{V}(\phi_{0})=:e^{i \beta_{n}  X^{\mu}(\phi_{0})}:$ of which the integral form is 
\begin{equation}
V_{\xi_{n}}(\phi_{0})=\int dt d\phi \sqrt{\gamma}\mathcal{V}(\phi_{0}).
\label{25}
\end{equation}
Here, $::$ means the normal ordering, and $\beta_{n}$ is the charge of $\xi_{n}$ which plays the role of space-time momentum along the $X^{\mu}$. $\gamma$ is the induced metric in the boundary. 
For a  physical state $\ket {\xi_{n}}$, it must be a BRST invariant state in BCFT at ghost number 1, satisfying \cite{Erler:2019vhl,Polchinski} 
\begin{equation}
Q\ket{\xi_{n}}=0,\;\;\; \ket{\xi_{n}} \sim \ket{\xi_{n}}+Q\ket\Lambda,\;\; \ket{\xi_{n}}=ghost \;number \;1,
\label{26}
\end{equation}
where $Q$ is the BRST operator and $\ket\Lambda$ is ghost number 0 state.
Recalling the equation of motion \eqref{22}, we find that 
it is natural to treat the normal fluctuation operator $O_{n}$ as the BRST operator $Q$. 

Next, we use the $Schr\ddot odinger$ representation for the excited worldsheet fields and consider Dirichlet boundary conditions of the probe string on both endpoints
\begin{equation}
X^{\mu}(\phi=0)=x^{\mu}, \;\; X^{\mu}(\phi=\pi)=x^{\mu}.
\label{27}
\end{equation}
Subsequently, since the background metric is conformally flat near the boundary, so we can expand $X^{\mu}(t,\phi)$ in (\ref{3}) as \cite{Maccaferri:2023wrg}
\begin{equation}
 X^{\mu}(\phi)= x^{\mu}(\phi)+\frac{( x^{\mu}_{\frac{\phi_{c}}{2}}- x^{\mu}_{\frac{-\phi_{c}}{2}})}{\pi}\phi+\sqrt{2\alpha^{'}}\sum_{n\ne0}\frac{1}{n}\alpha^{\mu}_{n}\sin(n\phi), \;\; 
\label{28}
\end{equation}
or more explicitly in the form 
\begin{eqnarray}
&& z(\phi)=z(\phi)+\sqrt{2\alpha^{'}} \sum_{n\ne 0}\frac{1}{n} \alpha^{\mu}_{n}\sin[n\phi], \notag \\
&& \phi=-\frac{\phi_{c}}{2}+\frac{\phi_{c}}{\pi} \phi+\sqrt{2\alpha^{'}} \sum_{n\ne 0}\frac{1}{n} \alpha^{\mu}_{n}\sin[n\phi],
\label{29}
\end{eqnarray}
and the vertex operator (\ref{24}) could be
 \begin{eqnarray}
 &&V_{\xi_{n}(z)}(\phi_{0})=e^{i\beta_{n}z_{*}} \times \exp[i\sqrt{2\alpha'}\beta_{n}\sum_{n\ne 0}\frac{\alpha^{\mu}_{n}}{n}\sin[n\phi_{0}]], \notag \\
 &&V_{\xi_{n}(\phi)}(\phi_{0})=e^{i\beta_{n}(\frac{\phi_{0}}{\pi}-\frac{1}{2})\phi_{c}} \times \exp[i\sqrt{2\alpha'}\beta_{n}\sum_{n\ne 0}\frac{\alpha^{\mu}_{n}}{n}\sin[n\phi_{0}]].
 \label{30} 
 \end{eqnarray}

Then we  define the overlap 
\begin{equation}
\Xi_{n}[X^{\mu}(\phi)]=\bra{X^{\mu}(\phi)}\Xi_{n}\rangle, 
\label{31}
\end{equation}
where $\Xi_{n}$ denotes the collection of excited fields on the worldsheet, depending on the curve $X^{\mu}$ in the background. It is noticed  that  $\Xi_{n}$ could also include the $b(\phi)$ and $c(\phi)$ fields, but here we do not consider this ghost sector. 
For the chosen point  $\phi_{0}$ in the boundary, we calculate the correlation of the vertex operator in the worldsheet as
\begin{equation}
\langle\Xi_{n}, \Xi_{n}^{'}\rangle=\langle (I\circ V_{\xi_{n}}(\phi_{0}))V_{\xi^{'}_{n}}(\phi_{0})\rangle_{worldsheet}.
\label{32}
\end{equation}
where $I$ denotes the Belavin-Polyakov-Zamolodchikov (BPZ) conjugation, $I(a)=-\frac{1}{a}$. That is to say, the $I \circ V_{\xi_{n}}(\phi_{0})$ corresponds to another point in the boundary. Thus,  (\ref{32}) contains two vertex operators, which are at $\phi_{0}$ and $-\frac{1}{\phi_{0}}$, respectively, and the formula of the path integral over the worldsheet is
\begin{equation}
\langle \Xi_{n},\Xi_{n}^{'}\rangle = \int[dX^{\mu}](I\circ V_{\xi_{n}}(\phi_{0}))V_{\xi^{'}_{n}}(\phi_{0})e^{-(S^{(0)}_{NG}+S^{(2)}_{NG})}.
\label{33}
\end{equation}
Then we factorize this integration into three parts, saying outside $(z>\left|z(\phi)\right|)$, on $(z=\left|z(\phi)\right|)$, and below $(z<\left|z(\phi)\right|)$ the probe string,
\begin{eqnarray}
  \langle \Xi_{n},\Xi_{n}^{'}\rangle = &&\int^{\frac{\phi_{c}}{2}}_{-\frac{\phi_{c}}{2}}[dX^{\mu}]_{z=\left|z(\phi)\right|}\int^{\frac{\phi_{c}}{2}}_{-\frac{\phi_{c}}{2}} [dX^{\mu}]_{z>\left|z(\phi)\right|} \notag \\
&&\times \int^{\frac{\phi_{c}}{2}}_{-\frac{\phi_{c}}{2}}[dX^{\mu}]_{z<\left|z(\phi)\right|}(I\circ V_{\xi_{n}}(\phi_{0}))V_{\xi^{'}_{n}}(\phi_{0})e^{-(S^{(0)}_{NG}+S^{(2)}_{NG})}.
\label{34}
\end{eqnarray}
by using the BPZ inverse, which takes the form 
\begin{eqnarray}
\langle \Xi_{n},\Xi_{n}^{'}\rangle &&=\int^{\frac{\phi_{c}}{2}}_{-\frac{\phi_{c}}{2}}[dX^{\mu}]_{z=\left|z(\phi)\right|}\left(\int^{\frac{\phi_{c}}{2}}_{-\frac{\phi_{c}}{2}}[dX^{\mu}]_{z<\left| z(\phi)\right|}V_{\xi_{n}}(\phi_{0})e^{-S^{(0)}_{NG}-S^{(2)}_{NG}} \right)  \notag \\ 
&&\times  \left(\int^{\frac{\phi_{c}}{2}}_{-\frac{\phi_{c}}{2}}[dX^{\mu}]_{z<\left|z(\phi)\right|}V_{\xi^{'}_{n}}(\phi_{0})e^{-S^{(0)}_{NG}-S^{(2)}_{NG}} \right). 
\label{35}
\end{eqnarray}

From (\ref{35}),  we can extract the induced string field on the worldsheet or the projection of excited fields of Hawking radiation on the worldsheet  as  
\begin{eqnarray}
\Xi_{n}[X^{\mu}(\phi)]&&=\int [dX^{\mu}]V_{\xi_{n}}(\phi_{0})e^{-S_{NG}^{(0)}-S_{NG}^{(2)}} \notag \\
&&= \int d(\delta X^{\mu}) V_{\xi_{n}}(\phi_{0})e^{-S^{(0)}_{NG}-S^{(2)}_{NG}} \notag \\
&&=\int n_{\mu}d\xi_{n}V_{\xi_{n}}(\phi_{0})e^{-S_{NG}^{(0)}-S_{NG}^{(2)}}.
\label{36}
\end{eqnarray}
In the second equality, we change the integration variable in term of the facts that $dX^{\mu}=dX^{0}+d(\delta X)$ and $dX^{0}$ is fixed. In the third equality,  we recall \eqref{8} and take $n_{\mu}$ as a constant vector along the worldsheet.  Then  considering that $\int d\xi e^{-S_{NG}^{(2)}}$ is a Gaussian integral, we shall further reduce the above formula into
\begin{equation}
\Xi_{n}[X^{\mu}]=n_{\mu}\int dt d\phi \sqrt{\frac{\pi}{T_{s}O_{n}}}\cdot V_{\xi_{n}}(\phi_{0})e^{-S^{0}_{NG}},
\label{37}
\end{equation}
which could be defined as the induced fields excited by Hawking radiation under the Schr$\ddot{o}$dinger functional representation.  As expected, $n_{\mu}$ appears here indicates that these fields are polarized and normal to the worldsheet. One interesting property we can read off from \eqref{37} is when the string tension $T_{s}$ approaches to zero, the fields will blow up. This could be understood as that the structure of string worldhseet will be changed  by the high Hawking temperature (Hagedorn temperature) when it goes near the horizon \cite{Giddings:1989xe,Bowick:1989us,Bagchi:2015nca}.

Before closing this section, we shall present some discussions on our results. Firstly, we argue that $\Xi_{n}[X^{\mu}]$ can be defined as the induced field excited by Hawking radiation based on the fact that  the fluctuation of worldsheet can be caused by Hawking radiation as the probe string is in black hole environment, even though the physical quantities related to Hawking radiation are not reflected in (\ref{37}). It is noticed that besides Hawking radiation, the fluctuations may also be triggered by other mechanisms, such as a scalar field on the worldsheet as a defect \cite{Garriga:1991tb}, but here we take more care of the worldsheet fluctuation itself , which means that fields $\Xi_{n}[X^{\mu}]$ are only curve-dependent.
Secondly, in our analysis, we see that under the second-order fluctuation on the action/complexity growth, the two-point correlation  emerge from the worldsheet's perturbation. Moreover, (\ref{35}) shows that the second order fluctuation of action growth of probe string indeed contributes to the correlation of vertex operators in the boundary conformal field theory in the form of functional path integrals. We argue that this may indicate a profound connection between complexity and the correlation function, which deserves further clarification.

\section{Conclusion}\label{summary}
There are plenty of schemes to define complexity in quantum field theory and conformal field theory. Relating with gravity, one also has two holographic versions of complexity. So further studies on complexity in a suitable framework will help us to further understand complexity in physics. In this work, we handle with complexity in string field theory and expect to shed light on connecting  complexity with correlation function in quantum field theory and holography.

We investigated the second-order quantum fluctuation of a probe string ending on the boundary of the BTZ black hole background. 
Since the string is in black hole environment, we assumed that 
the unique normal fluctuation  on the string is introduced by the Hawking radiation. Considering that the  fluctuation can also affect the complexity or action growth of the string and will bring in a high-order correction on the Nambu-Goto action. So we  calculate the second-order correction on the Nambu-Goto action and obtain the normal fluctuation operator.

Moreover, in the SFT/BCFT framework, by treating the excited fields on the worldsheet as a single object, and further defining them as excited string field, we found that the normal fluctuation operator also can be viewed as a BRST operator. 
Then we calculated the correlation function of vertex operators which is given in the form of functional path integrals. This corresponds to the correlation function of the excited fields on the worldsheet.  We then extracted the induced field excited by Hawking radiation under the Schr$\ddot o$dinger functional representation. Our study shows  that under the second-order fluctuation on the action/complexity growth, the two-point correlation can emerge from the worldsheet's perturbation. This indicates that somehow we may explain the complexity  as scattering amplitude, which could be an interesting direct, for example, one should clarify the definition of the minimal path in the scattering amplitude, and so on.  We hope to further study the related topics in the future.

\acknowledgments{}

We appreciate Profs. Rui-Hong Yue, Ya-Peng Hu  and Jian-Pin Wu for helpful and intriguing discussions. Moreover, We are grateful to Dr. Guo-Yang Fu and Dr. Kang Zhou for their constructive advice. This work is supported by National Natural Science Foundation of China (NSFC) under grant Nos. 12247170, 12175105, Top- notch Academic Programs Project of Jiangsu Higher Education Institutions (TAPP).

\bibliographystyle{utphys}
\bibliography{ref}

\providecommand{\href}[2]{#2}\begingroup\raggedright\begin{thebibliography}{10}

\bibitem{Susskind:1994vu}
L.~Susskind, ``{The World as a hologram},''
  \href{http://dx.doi.org/10.1063/1.531249}{{\em J. Math. Phys.} {\bfseries 36}
  (1995) 6377--6396}, \href{http://arxiv.org/abs/hep-th/9409089}{{\ttfamily
  arXiv:hep-th/9409089}}.

\bibitem{tHooft:1999rgb}
G.~'t~Hooft, ``{The Holographic principle: Opening lecture},''
  \href{http://dx.doi.org/10.1142/9789812811585_0005}{{\em Subnucl. Ser.}
  {\bfseries 37} (2001) 72--100},
  \href{http://arxiv.org/abs/hep-th/0003004}{{\ttfamily arXiv:hep-th/0003004}}.

\bibitem{Maldacena:1997re}
J.~M. Maldacena, ``{The Large N limit of superconformal field theories and
  supergravity},'' \href{http://dx.doi.org/10.4310/ATMP.1998.v2.n2.a1}{{\em
  Adv. Theor. Math. Phys.} {\bfseries 2} (1998) 231--252},
  \href{http://arxiv.org/abs/hep-th/9711200}{{\ttfamily arXiv:hep-th/9711200}}.

\bibitem{Gubser:1998bc}
S.~S. Gubser, I.~R. Klebanov, and A.~M. Polyakov, ``{Gauge theory correlators
  from noncritical string theory},''
  \href{http://dx.doi.org/10.1016/S0370-2693(98)00377-3}{{\em Phys. Lett. B}
  {\bfseries 428} (1998) 105--114},
  \href{http://arxiv.org/abs/hep-th/9802109}{{\ttfamily arXiv:hep-th/9802109}}.

\bibitem{Witten:1998qj}
E.~Witten, ``{Anti-de Sitter space and holography},''
  \href{http://dx.doi.org/10.4310/ATMP.1998.v2.n2.a2}{{\em Adv. Theor. Math.
  Phys.} {\bfseries 2} (1998) 253--291},
  \href{http://arxiv.org/abs/hep-th/9802150}{{\ttfamily arXiv:hep-th/9802150}}.

\bibitem{Osborne_2012}
T.~J. Osborne, ``Hamiltonian complexity,''
  \href{http://dx.doi.org/10.1088/0034-4885/75/2/022001}{{\em Reports on
  Progress in Physics} {\bfseries 75} no.~2, (Jan, 2012) 022001}.

\bibitem{TCS-066}
S.~Gharibian, Y.~Huang, Z.~Landau, and S.~W. Shin, ``Quantum hamiltonian
  complexity,'' \href{http://dx.doi.org/10.1561/0400000066}{{\em Foundations
  and Trends® in Theoretical Computer Science} {\bfseries 10} no.~3, (2015)
  159--282}.

\bibitem{Dvali000}
G.~Dvali, C.~Gomez, D.~Lüst, Y.~Omar, and B.~Richter, ``Universality of black
  hole quantum computing,''
  \href{http://dx.doi.org/https://doi.org/10.1002/prop.201600111}{{\em
  Fortschritte der Physik} {\bfseries 65} no.~1, (2017) 1600111}.

\bibitem{PhysRevA.94.040302}
B.~Swingle, G.~Bentsen, M.~Schleier-Smith, and P.~Hayden, ``Measuring the
  scrambling of quantum information,''
  \href{http://dx.doi.org/10.1103/PhysRevA.94.040302}{{\em Phys. Rev. A}
  {\bfseries 94} (Oct, 2016) 040302}.

\bibitem{PhysRevD.96.126001}
K.~Hashimoto, N.~Iizuka, and S.~Sugishita, ``Time evolution of complexity in
  abelian gauge theories,''
  \href{http://dx.doi.org/10.1103/PhysRevD.96.126001}{{\em Phys. Rev. D}
  {\bfseries 96} (Dec, 2017) 126001}.

\bibitem{Watrous:2008}
J.Watrous, ``{Quantum computational complexity},''
  \href{http://arxiv.org/abs/0804.3401}{{\ttfamily arXiv:0804.3401
  [quant-ph]}}.

\bibitem{Bao:2018ira}
N.~Bao and J.~Liu, ``{Quantum complexity and the virial theorem},''
  \href{http://dx.doi.org/10.1007/JHEP08(2018)144}{{\em JHEP} {\bfseries 08}
  (2018) 144}, \href{http://arxiv.org/abs/1804.03242}{{\ttfamily
  arXiv:1804.03242 [hep-th]}}.

\bibitem{Vanchurin:2016met}
V.~Vanchurin, ``{Dual Field Theories of Quantum Computation},''
  \href{http://dx.doi.org/10.1007/JHEP06(2016)001}{{\em JHEP} {\bfseries 06}
  (2016) 001}, \href{http://arxiv.org/abs/1603.07982}{{\ttfamily
  arXiv:1603.07982 [hep-th]}}.

\bibitem{Chapman:2017rqy}
S.~Chapman, M.~P. Heller, H.~Marrochio, and F.~Pastawski, ``{Toward a
  Definition of Complexity for Quantum Field Theory States},''
  \href{http://dx.doi.org/10.1103/PhysRevLett.120.121602}{{\em Phys. Rev.
  Lett.} {\bfseries 120} no.~12, (2018) 121602},
  \href{http://arxiv.org/abs/1707.08582}{{\ttfamily arXiv:1707.08582
  [hep-th]}}.

\bibitem{Molina-Vilaplana:2018sfn}
J.~Molina-Vilaplana and A.~Del~Campo, ``{Complexity Functionals and Complexity
  Growth Limits in Continuous MERA Circuits},''
  \href{http://dx.doi.org/10.1007/JHEP08(2018)012}{{\em JHEP} {\bfseries 08}
  (2018) 012}, \href{http://arxiv.org/abs/1803.02356}{{\ttfamily
  arXiv:1803.02356 [hep-th]}}.

\bibitem{Bhattacharyya:2018wym}
A.~Bhattacharyya, P.~Caputa, S.~R. Das, N.~Kundu, M.~Miyaji, and T.~Takayanagi,
  ``{Path-Integral Complexity for Perturbed CFTs},''
  \href{http://dx.doi.org/10.1007/JHEP07(2018)086}{{\em JHEP} {\bfseries 07}
  (2018) 086}, \href{http://arxiv.org/abs/1804.01999}{{\ttfamily
  arXiv:1804.01999 [hep-th]}}.

\bibitem{Nielsen:2005mkt}
M.~A. Nielsen, ``{A geometric approach to quantum circuit lower bounds},''
  \href{http://dx.doi.org/10.26421/QIC6.3-2}{{\em Quant. Inf. Comput.}
  {\bfseries 6} no.~3, (2006) 213--262},
  \href{http://arxiv.org/abs/quant-ph/0502070}{{\ttfamily
  arXiv:quant-ph/0502070}}.

\bibitem{doi:10.1126/science.1121541}
M.~A. Nielsen, M.~R. Dowling, M.~Gu, and A.~C. Doherty, ``Quantum computation
  as geometry,'' \href{http://dx.doi.org/10.1126/science.1121541}{{\em Science}
  {\bfseries 311} no.~5764, (2006) 1133--1135}.

\bibitem{Jefferson:2017sdb}
R.~Jefferson and R.~C. Myers, ``{Circuit complexity in quantum field theory},''
  \href{http://dx.doi.org/10.1007/JHEP10(2017)107}{{\em JHEP} {\bfseries 10}
  (2017) 107}, \href{http://arxiv.org/abs/1707.08570}{{\ttfamily
  arXiv:1707.08570 [hep-th]}}.

\bibitem{Yang:2018nda}
R.-Q. Yang, Y.-S. An, C.~Niu, C.-Y. Zhang, and K.-Y. Kim, ``{Principles and
  symmetries of complexity in quantum field theory},''
  \href{http://dx.doi.org/10.1140/epjc/s10052-019-6600-3}{{\em Eur. Phys. J. C}
  {\bfseries 79} no.~2, (2019) 109},
  \href{http://arxiv.org/abs/1803.01797}{{\ttfamily arXiv:1803.01797
  [hep-th]}}.

\bibitem{Bhattacharyya:2018bbv}
A.~Bhattacharyya, A.~Shekar, and A.~Sinha, ``{Circuit complexity in interacting
  QFTs and RG flows},'' \href{http://dx.doi.org/10.1007/JHEP10(2018)140}{{\em
  JHEP} {\bfseries 10} (2018) 140},
  \href{http://arxiv.org/abs/1808.03105}{{\ttfamily arXiv:1808.03105
  [hep-th]}}.

\bibitem{Bhattacharyya:2019kvj}
A.~Bhattacharyya, P.~Nandy, and A.~Sinha, ``{Renormalized Circuit
  Complexity},'' \href{http://dx.doi.org/10.1103/PhysRevLett.124.101602}{{\em
  Phys. Rev. Lett.} {\bfseries 124} no.~10, (2020) 101602},
  \href{http://arxiv.org/abs/1907.08223}{{\ttfamily arXiv:1907.08223
  [hep-th]}}.

\bibitem{Camargo:2022wkd}
H.~A. Camargo, P.~Caputa, and P.~Nandy, ``{Q-curvature and path integral
  complexity},'' \href{http://dx.doi.org/10.1007/JHEP04(2022)081}{{\em JHEP}
  {\bfseries 04} (2022) 081}, \href{http://arxiv.org/abs/2201.00562}{{\ttfamily
  arXiv:2201.00562 [hep-th]}}. [Erratum: JHEP 10, 038 (2023)].

\bibitem{Adhikari:2022whf}
K.~Adhikari, S.~Choudhury, and A.~Roy, ``{Krylov Complexity in Quantum Field
  Theory},'' \href{http://dx.doi.org/10.1016/j.nuclphysb.2023.116263}{{\em
  Nucl. Phys. B} {\bfseries 993} (2023) 116263},
  \href{http://arxiv.org/abs/2204.02250}{{\ttfamily arXiv:2204.02250
  [hep-th]}}.

\bibitem{Adhikari:2021pvv}
K.~Adhikari, S.~Choudhury, S.~Chowdhury, K.~Shirish, and A.~Swain, ``{Circuit
  complexity as a novel probe of quantum entanglement: A study with black hole
  gas in arbitrary dimensions},''
  \href{http://dx.doi.org/10.1103/PhysRevD.104.065002}{{\em Phys. Rev. D}
  {\bfseries 104} no.~6, (2021) 065002},
  \href{http://arxiv.org/abs/2104.13940}{{\ttfamily arXiv:2104.13940
  [hep-th]}}.

\bibitem{Adhikari:2022oxr}
K.~Adhikari and S.~Choudhury, ``{Cosmological Krylov Complexity},''
  \href{http://dx.doi.org/10.1002/prop.202200126}{{\em Fortsch. Phys.}
  {\bfseries 70} no.~12, (2022) 2200126},
  \href{http://arxiv.org/abs/2203.14330}{{\ttfamily arXiv:2203.14330
  [hep-th]}}.

\bibitem{Chapman:2016hwi}
S.~Chapman, H.~Marrochio, and R.~C. Myers, ``{Complexity of Formation in
  Holography},'' \href{http://dx.doi.org/10.1007/JHEP01(2017)062}{{\em JHEP}
  {\bfseries 01} (2017) 062}, \href{http://arxiv.org/abs/1610.08063}{{\ttfamily
  arXiv:1610.08063 [hep-th]}}.

\bibitem{Brown:2015bva}
A.~R. Brown, D.~A. Roberts, L.~Susskind, B.~Swingle, and Y.~Zhao,
  ``{Holographic Complexity Equals Bulk Action?},''
  \href{http://dx.doi.org/10.1103/PhysRevLett.116.191301}{{\em Phys. Rev.
  Lett.} {\bfseries 116} no.~19, (2016) 191301},
  \href{http://arxiv.org/abs/1509.07876}{{\ttfamily arXiv:1509.07876
  [hep-th]}}.

\bibitem{Brown:2015lvg}
A.~R. Brown, D.~A. Roberts, L.~Susskind, B.~Swingle, and Y.~Zhao,
  ``{Complexity, action, and black holes},''
  \href{http://dx.doi.org/10.1103/PhysRevD.93.086006}{{\em Phys. Rev. D}
  {\bfseries 93} no.~8, (2016) 086006},
  \href{http://arxiv.org/abs/1512.04993}{{\ttfamily arXiv:1512.04993
  [hep-th]}}.

\bibitem{Pan:2016ecg}
W.-J. Pan and Y.-C. Huang, ``{Holographic complexity and action growth in
  massive gravities},''
  \href{http://dx.doi.org/10.1103/PhysRevD.95.126013}{{\em Phys. Rev. D}
  {\bfseries 95} no.~12, (2017) 126013},
  \href{http://arxiv.org/abs/1612.03627}{{\ttfamily arXiv:1612.03627
  [hep-th]}}.

\bibitem{Guo:2017rul}
W.-D. Guo, S.-W. Wei, Y.-Y. Li, and Y.-X. Liu, ``{Complexity growth rates for
  AdS black holes in massive gravity and $f(R)$ gravity},''
  \href{http://dx.doi.org/10.1140/epjc/s10052-017-5466-5}{{\em Eur. Phys. J. C}
  {\bfseries 77} no.~12, (2017) 904},
  \href{http://arxiv.org/abs/1703.10468}{{\ttfamily arXiv:1703.10468 [gr-qc]}}.

\bibitem{Momeni:2016ekm}
D.~Momeni, S.~A. Hosseini~Mansoori, and R.~Myrzakulov, ``{Holographic
  Complexity in Gauge/String Superconductors},''
  \href{http://dx.doi.org/10.1016/j.physletb.2016.03.031}{{\em Phys. Lett. B}
  {\bfseries 756} (2016) 354--357},
  \href{http://arxiv.org/abs/1601.03011}{{\ttfamily arXiv:1601.03011
  [hep-th]}}.

\bibitem{Tao:2017fsy}
J.~Tao, P.~Wang, and H.~Yang, ``{Testing holographic conjectures of complexity
  with Born\textendash{}Infeld black holes},''
  \href{http://dx.doi.org/10.1140/epjc/s10052-017-5395-3}{{\em Eur. Phys. J. C}
  {\bfseries 77} no.~12, (2017) 817},
  \href{http://arxiv.org/abs/1703.06297}{{\ttfamily arXiv:1703.06297
  [hep-th]}}.

\bibitem{Alishahiha:2017hwg}
M.~Alishahiha, A.~Faraji~Astaneh, A.~Naseh, and M.~H. Vahidinia, ``{On
  complexity for F(R) and critical gravity},''
  \href{http://dx.doi.org/10.1007/JHEP05(2017)009}{{\em JHEP} {\bfseries 05}
  (2017) 009}, \href{http://arxiv.org/abs/1702.06796}{{\ttfamily
  arXiv:1702.06796 [hep-th]}}.

\bibitem{Reynolds:2017lwq}
A.~Reynolds and S.~F. Ross, ``{Complexity in de Sitter Space},''
  \href{http://dx.doi.org/10.1088/1361-6382/aa8122}{{\em Class. Quant. Grav.}
  {\bfseries 34} no.~17, (2017) 175013},
  \href{http://arxiv.org/abs/1706.03788}{{\ttfamily arXiv:1706.03788
  [hep-th]}}.

\bibitem{Qaemmaqami:2017lzs}
M.~M. Qaemmaqami, ``{Complexity growth in minimal massive 3D gravity},''
  \href{http://dx.doi.org/10.1103/PhysRevD.97.026006}{{\em Phys. Rev. D}
  {\bfseries 97} no.~2, (2018) 026006},
  \href{http://arxiv.org/abs/1709.05894}{{\ttfamily arXiv:1709.05894
  [hep-th]}}.

\bibitem{Sebastiani:2017rxr}
L.~Sebastiani, L.~Vanzo, and S.~Zerbini, ``{Action growth for black holes in
  modified gravity},'' \href{http://dx.doi.org/10.1103/PhysRevD.97.044009}{{\em
  Phys. Rev. D} {\bfseries 97} no.~4, (2018) 044009},
  \href{http://arxiv.org/abs/1710.05686}{{\ttfamily arXiv:1710.05686
  [hep-th]}}.

\bibitem{Couch:2017yil}
J.~Couch, S.~Eccles, W.~Fischler, and M.-L. Xiao, ``{Holographic complexity and
  noncommutative gauge theory},''
  \href{http://dx.doi.org/10.1007/JHEP03(2018)108}{{\em JHEP} {\bfseries 03}
  (2018) 108}, \href{http://arxiv.org/abs/1710.07833}{{\ttfamily
  arXiv:1710.07833 [hep-th]}}.

\bibitem{Swingle:2017zcd}
B.~Swingle and Y.~Wang, ``{Holographic Complexity of Einstein-Maxwell-Dilaton
  Gravity},'' \href{http://dx.doi.org/10.1007/JHEP09(2018)106}{{\em JHEP}
  {\bfseries 09} (2018) 106}, \href{http://arxiv.org/abs/1712.09826}{{\ttfamily
  arXiv:1712.09826 [hep-th]}}.

\bibitem{Cano:2018aqi}
P.~A. Cano, R.~A. Hennigar, and H.~Marrochio, ``{Complexity Growth Rate in
  Lovelock Gravity},''
  \href{http://dx.doi.org/10.1103/PhysRevLett.121.121602}{{\em Phys. Rev.
  Lett.} {\bfseries 121} no.~12, (2018) 121602},
  \href{http://arxiv.org/abs/1803.02795}{{\ttfamily arXiv:1803.02795
  [hep-th]}}.

\bibitem{Chapman:2018dem}
S.~Chapman, H.~Marrochio, and R.~C. Myers, ``{Holographic complexity in Vaidya
  spacetimes. Part I},'' \href{http://dx.doi.org/10.1007/JHEP06(2018)046}{{\em
  JHEP} {\bfseries 06} (2018) 046},
  \href{http://arxiv.org/abs/1804.07410}{{\ttfamily arXiv:1804.07410
  [hep-th]}}.

\bibitem{Chapman:2018lsv}
S.~Chapman, H.~Marrochio, and R.~C. Myers, ``{Holographic complexity in Vaidya
  spacetimes. Part II},'' \href{http://dx.doi.org/10.1007/JHEP06(2018)114}{{\em
  JHEP} {\bfseries 06} (2018) 114},
  \href{http://arxiv.org/abs/1805.07262}{{\ttfamily arXiv:1805.07262
  [hep-th]}}.

\bibitem{Auzzi:2018pbc}
R.~Auzzi, S.~Baiguera, M.~Grassi, G.~Nardelli, and N.~Zenoni, ``{Complexity and
  action for warped AdS black holes},''
  \href{http://dx.doi.org/10.1007/JHEP09(2018)013}{{\em JHEP} {\bfseries 09}
  (2018) 013}, \href{http://arxiv.org/abs/1806.06216}{{\ttfamily
  arXiv:1806.06216 [hep-th]}}.

\bibitem{Yaraie:2018hwz}
E.~Yaraie, H.~Ghaffarnejad, and M.~Farsam, ``{Complexity growth and shock wave
  geometry in AdS-Maxwell-power-Yang\textendash{}Mills theory},''
  \href{http://dx.doi.org/10.1140/epjc/s10052-018-6456-y}{{\em Eur. Phys. J. C}
  {\bfseries 78} no.~11, (2018) 967},
  \href{http://arxiv.org/abs/1806.07242}{{\ttfamily arXiv:1806.07242 [gr-qc]}}.

\bibitem{Alishahiha:2018tep}
M.~Alishahiha, A.~Faraji~Astaneh, M.~R. Mohammadi~Mozaffar, and A.~Mollabashi,
  ``{Complexity Growth with Lifshitz Scaling and Hyperscaling Violation},''
  \href{http://dx.doi.org/10.1007/JHEP07(2018)042}{{\em JHEP} {\bfseries 07}
  (2018) 042}, \href{http://arxiv.org/abs/1802.06740}{{\ttfamily
  arXiv:1802.06740 [hep-th]}}.

\bibitem{An:2018xhv}
Y.-S. An and R.-H. Peng, ``{Effect of the dilaton on holographic complexity
  growth},'' \href{http://dx.doi.org/10.1103/PhysRevD.97.066022}{{\em Phys.
  Rev. D} {\bfseries 97} no.~6, (2018) 066022},
  \href{http://arxiv.org/abs/1801.03638}{{\ttfamily arXiv:1801.03638
  [hep-th]}}.

\bibitem{Cai:2016xho}
R.-G. Cai, S.-M. Ruan, S.-J. Wang, R.-Q. Yang, and R.-H. Peng, ``{Action growth
  for AdS black holes},'' \href{http://dx.doi.org/10.1007/JHEP09(2016)161}{{\em
  JHEP} {\bfseries 09} (2016) 161},
  \href{http://arxiv.org/abs/1606.08307}{{\ttfamily arXiv:1606.08307 [gr-qc]}}.

\bibitem{Ghodsi:2020qqb}
A.~Ghodsi, S.~Qolibikloo, and S.~Karimi, ``{Holographic complexity in general
  quadratic curvature theory of gravity},''
  \href{http://dx.doi.org/10.1140/epjc/s10052-020-08503-9}{{\em Eur. Phys. J.
  C} {\bfseries 80} no.~10, (2020) 920},
  \href{http://arxiv.org/abs/2005.08989}{{\ttfamily arXiv:2005.08989
  [hep-th]}}.

\bibitem{Frassino:2019fgr}
A.~M. Frassino, R.~B. Mann, and J.~R. Mureika, ``{Extended Thermodynamics and
  Complexity in Gravitational Chern-Simons Theory},''
  \href{http://dx.doi.org/10.1007/JHEP11(2019)112}{{\em JHEP} {\bfseries 11}
  (2019) 112}, \href{http://arxiv.org/abs/1906.07190}{{\ttfamily
  arXiv:1906.07190 [gr-qc]}}.

\bibitem{Abad:2017cgl}
F.~J.~G. Abad, M.~Kulaxizi, and A.~Parnachev, ``{On Complexity of Holographic
  Flavors},'' \href{http://dx.doi.org/10.1007/JHEP01(2018)127}{{\em JHEP}
  {\bfseries 01} (2018) 127}, \href{http://arxiv.org/abs/1705.08424}{{\ttfamily
  arXiv:1705.08424 [hep-th]}}.

\bibitem{Santos:2022lxj}
F.~F. Santos, O.~Sokoliuk, and A.~Baransky, ``{Holographic Complexity of
  Braneworld in Horndeski Gravity},''
  \href{http://dx.doi.org/10.1002/prop.202200141}{{\em Fortsch. Phys.}
  {\bfseries 71} no.~2-3, (2023) 2200141},
  \href{http://arxiv.org/abs/2210.11596}{{\ttfamily arXiv:2210.11596
  [hep-th]}}.

\bibitem{Ageev:2014nva}
D.~S. Ageev and I.~Y. Aref'eva, ``{Holography and nonlocal operators for the
  BTZ black hole with nonzero angular momentum},''
  \href{http://dx.doi.org/10.1007/s11232-014-0186-6}{{\em Theor. Math. Phys.}
  {\bfseries 180} (2014) 881--893},
  \href{http://arxiv.org/abs/1402.6937}{{\ttfamily arXiv:1402.6937 [hep-th]}}.

\bibitem{Zhou:2021vsm}
Y.-T. Zhou, X.-M. Kuang, and J.-P. Wu, ``{Complexity growth of massive black
  hole with a probe string},''
  \href{http://dx.doi.org/10.1140/epjc/s10052-021-09563-1}{{\em Eur. Phys. J.
  C} {\bfseries 81} no.~8, (2021) 768},
  \href{http://arxiv.org/abs/2104.12998}{{\ttfamily arXiv:2104.12998
  [hep-th]}}.

\bibitem{Zhou:2023nza}
Y.-T. Zhou, ``{Complexity growth of BTZ black hole in massive gravity with a
  null string},'' \href{http://dx.doi.org/10.1140/epjc/s10052-023-12260-w}{{\em
  Eur. Phys. J. C} {\bfseries 83} no.~12, (2023) 1109},
  \href{http://arxiv.org/abs/2302.10565}{{\ttfamily arXiv:2302.10565
  [hep-th]}}.

\bibitem{Nagasaki:2018csh}
K.~Nagasaki, ``{Complexity growth of rotating black holes with a probe
  string},'' \href{http://dx.doi.org/10.1103/PhysRevD.98.126014}{{\em Phys.
  Rev. D} {\bfseries 98} no.~12, (2018) 126014},
  \href{http://arxiv.org/abs/1807.01088}{{\ttfamily arXiv:1807.01088
  [hep-th]}}.

\bibitem{Nagasaki:2019icm}
K.~Nagasaki, ``{Complexity growth for topological black holes by holographic
  method},'' \href{http://dx.doi.org/10.1142/S0217751X20501523}{{\em Int. J.
  Mod. Phys. A} {\bfseries 35} no.~25, (2020) 2050152},
  \href{http://arxiv.org/abs/1912.03567}{{\ttfamily arXiv:1912.03567
  [hep-th]}}.

\bibitem{Bravo-Gaete:2020lzs}
M.~Bravo-Gaete and F.~F. Santos, ``{Complexity of four-dimensional hairy
  anti-de-Sitter black holes with a rotating string and shear viscosity in
  generalized scalar\textendash{}tensor theories},''
  \href{http://dx.doi.org/10.1140/epjc/s10052-022-10064-y}{{\em Eur. Phys. J.
  C} {\bfseries 82} no.~2, (2022) 101},
  \href{http://arxiv.org/abs/2010.10942}{{\ttfamily arXiv:2010.10942
  [hep-th]}}.

\bibitem{Santos:2020xox}
F.~F. Santos, ``{Rotating black hole with a probe string in Horndeski
  Gravity},'' \href{http://dx.doi.org/10.1140/epjp/s13360-020-00805-x}{{\em
  Eur. Phys. J. Plus} {\bfseries 135} no.~10, (2020) 810},
  \href{http://arxiv.org/abs/2005.10983}{{\ttfamily arXiv:2005.10983
  [hep-th]}}.

\bibitem{Nagasaki:2021ldz}
K.~Nagasaki, ``{Probe strings on anti-de Sitter accelerating black holes},''
  \href{http://dx.doi.org/10.1093/ptep/ptac038}{{\em PTEP} {\bfseries 2022}
  no.~4, (2022) 043B02}, \href{http://arxiv.org/abs/2108.05429}{{\ttfamily
  arXiv:2108.05429 [hep-th]}}.

\bibitem{Nagasaki:2022lll}
K.~Nagasaki, ``{Effects of the acceleration on holographic complexity},''
  \href{http://dx.doi.org/10.1142/S0217751X23500276}{{\em Int. J. Mod. Phys. A}
  {\bfseries 38} no.~04n05, (2023) 2350027},
  \href{http://arxiv.org/abs/2205.00196}{{\ttfamily arXiv:2205.00196
  [hep-th]}}.

\bibitem{Susskind:2021esx}
L.~Susskind, ``{Entanglement and Chaos in De Sitter Space Holography: An SYK
  Example},'' \href{http://dx.doi.org/10.22128/jhap.2021.455.1005}{{\em JHAP}
  {\bfseries 1} no.~1, (2021) 1--22},
  \href{http://arxiv.org/abs/2109.14104}{{\ttfamily arXiv:2109.14104
  [hep-th]}}.

\bibitem{Jorstad:2022mls}
E.~J\o{}rstad, R.~C. Myers, and S.-M. Ruan, ``{Holographic complexity in
  dS$_{d+1}$},'' \href{http://dx.doi.org/10.1007/JHEP05(2022)119}{{\em JHEP}
  {\bfseries 05} (2022) 119}, \href{http://arxiv.org/abs/2202.10684}{{\ttfamily
  arXiv:2202.10684 [hep-th]}}.

\bibitem{Santos:2023eqp}
F.~F. Santos, B.~Pourhassan, and E.~N. Saridakis, ``{de Sitter Versus Anti-de
  Sitter in Horndeski-Like Gravity},''
  \href{http://dx.doi.org/10.1002/prop.202300228}{{\em Fortsch. Phys.}
  {\bfseries 72} no.~3, (2024) 2300228},
  \href{http://arxiv.org/abs/2305.05794}{{\ttfamily arXiv:2305.05794
  [hep-th]}}.

\bibitem{Brown:2017jil}
A.~R. Brown and L.~Susskind, ``{Second law of quantum complexity},''
  \href{http://dx.doi.org/10.1103/PhysRevD.97.086015}{{\em Phys. Rev. D}
  {\bfseries 97} no.~8, (2018) 086015},
  \href{http://arxiv.org/abs/1701.01107}{{\ttfamily arXiv:1701.01107
  [hep-th]}}.

\bibitem{Santos:2024zoh}
F.~F. Santos and H.~Boschi-Filho, ``{Holographic complexity and residual
  entropy of a rotating charged BTZ black hole within Horndeski gravity},''
  \href{http://arxiv.org/abs/2407.10004}{{\ttfamily arXiv:2407.10004
  [hep-th]}}.

\bibitem{deBoer:2008gu}
J.~de~Boer, V.~E. Hubeny, M.~Rangamani, and M.~Shigemori, ``{Brownian motion in
  AdS/CFT},'' \href{http://dx.doi.org/10.1088/1126-6708/2009/07/094}{{\em JHEP}
  {\bfseries 07} (2009) 094}, \href{http://arxiv.org/abs/0812.5112}{{\ttfamily
  arXiv:0812.5112 [hep-th]}}.

\bibitem{Atmaja:2010uu}
A.~N. Atmaja, J.~de~Boer, and M.~Shigemori, ``{Holographic Brownian Motion and
  Time Scales in Strongly Coupled Plasmas},''
  \href{http://dx.doi.org/10.1016/j.nuclphysb.2013.12.018}{{\em Nucl. Phys. B}
  {\bfseries 880} (2014) 23--75},
  \href{http://arxiv.org/abs/1002.2429}{{\ttfamily arXiv:1002.2429 [hep-th]}}.

\bibitem{Lawrence:1993sg}
A.~E. Lawrence and E.~J. Martinec, ``{Black hole evaporation along macroscopic
  strings},'' \href{http://dx.doi.org/10.1103/PhysRevD.50.2680}{{\em Phys. Rev.
  D} {\bfseries 50} (1994) 2680--2691},
  \href{http://arxiv.org/abs/hep-th/9312127}{{\ttfamily arXiv:hep-th/9312127}}.

\bibitem{Fuchs:2008cc}
E.~Fuchs and M.~Kroyter, ``{Analytical Solutions of Open String Field
  Theory},'' \href{http://dx.doi.org/10.1016/j.physrep.2011.01.003}{{\em Phys.
  Rept.} {\bfseries 502} (2011) 89--149},
  \href{http://arxiv.org/abs/0807.4722}{{\ttfamily arXiv:0807.4722 [hep-th]}}.

\bibitem{Taylor:2003gn}
W.~Taylor and B.~Zwiebach,
  \href{http://dx.doi.org/10.1142/9789812702821_0012}{``{D-branes, tachyons,
  and string field theory},''} pp.~641--759.
\newblock 10, 2003.
\newblock \href{http://arxiv.org/abs/hep-th/0311017}{{\ttfamily
  arXiv:hep-th/0311017}}.

\bibitem{Rastelli:2005mz}
L.~Rastelli, ``{String field theory},''
\newblock 9, 2005.
\newblock \href{http://arxiv.org/abs/hep-th/0509129}{{\ttfamily
  arXiv:hep-th/0509129}}.

\bibitem{Taylor:2006ye}
W.~Taylor, ``{String field theory},''
  \href{http://arxiv.org/abs/hep-th/0605202}{{\ttfamily arXiv:hep-th/0605202}}.

\bibitem{Kiermaier:2008qu}
M.~Kiermaier, Y.~Okawa, and B.~Zwiebach, ``{The boundary state from open string
  fields},'' \href{http://arxiv.org/abs/0810.1737}{{\ttfamily arXiv:0810.1737
  [hep-th]}}.

\bibitem{Kudrna:2012re}
M.~Kudrna, C.~Maccaferri, and M.~Schnabl, ``{Boundary State from Ellwood
  Invariants},'' \href{http://dx.doi.org/10.1007/JHEP07(2013)033}{{\em JHEP}
  {\bfseries 07} (2013) 033}, \href{http://arxiv.org/abs/1207.4785}{{\ttfamily
  arXiv:1207.4785 [hep-th]}}.

\bibitem{Santos:2021orr}
F.~F. Santos, E.~F. Capossoli, and H.~Boschi-Filho, ``{AdS/BCFT correspondence
  and BTZ black hole thermodynamics within Horndeski gravity},''
  \href{http://dx.doi.org/10.1103/PhysRevD.104.066014}{{\em Phys. Rev. D}
  {\bfseries 104} no.~6, (2021) 066014},
  \href{http://arxiv.org/abs/2105.03802}{{\ttfamily arXiv:2105.03802
  [hep-th]}}.

\bibitem{PhysRevLett.69.1849}
M.~Ba\~nados, C.~Teitelboim, and J.~Zanelli, ``Black hole in three-dimensional
  spacetime,'' \href{http://dx.doi.org/10.1103/PhysRevLett.69.1849}{{\em Phys.
  Rev. Lett.} {\bfseries 69} (Sep, 1992) 1849--1851}.

\bibitem{PhysRevD.48.1506}
M.~Ba\~nados, M.~Henneaux, C.~Teitelboim, and J.~Zanelli, ``Geometry of the 2+1
  black hole,'' \href{http://dx.doi.org/10.1103/PhysRevD.48.1506}{{\em Phys.
  Rev. D} {\bfseries 48} (Aug, 1993) 1506--1525}.

\bibitem{Zhou:2019jlh}
Y.-T. Zhou, M.~Ghodrati, X.-M. Kuang, and J.-P. Wu, ``{Evolutions of
  entanglement and complexity after a thermal quench in massive gravity
  theory},'' \href{http://dx.doi.org/10.1103/PhysRevD.100.066003}{{\em Phys.
  Rev. D} {\bfseries 100} no.~6, (2019) 066003},
  \href{http://arxiv.org/abs/1907.08453}{{\ttfamily arXiv:1907.08453
  [hep-th]}}.

\bibitem{Kinar:1999xu}
Y.~Kinar, E.~Schreiber, J.~Sonnenschein, and N.~Weiss, ``{Quantum fluctuations
  of Wilson loops from string models},''
  \href{http://dx.doi.org/10.1016/S0550-3213(00)00238-8}{{\em Nucl. Phys. B}
  {\bfseries 583} (2000) 76--104},
  \href{http://arxiv.org/abs/hep-th/9911123}{{\ttfamily arXiv:hep-th/9911123}}.

\bibitem{Erler:2019vhl}
T.~Erler, ``{Four lectures on analytic solutions in open string field
  theory},'' \href{http://dx.doi.org/10.1016/j.physrep.2022.06.004}{{\em Phys.
  Rept.} {\bfseries 980} (2022) 1--95},
  \href{http://arxiv.org/abs/1912.00521}{{\ttfamily arXiv:1912.00521
  [hep-th]}}.

\bibitem{Polchinski}
J.~Polchinski, {\em String theory, Vol. 1 \& 2, Cambridge University Press}.

\bibitem{Maccaferri:2023wrg}
C.~Maccaferri, F.~Marino, and B.~Valsesia, ``{Introduction to String Theory},''
  \href{http://arxiv.org/abs/2311.18111}{{\ttfamily arXiv:2311.18111
  [hep-th]}}.

\bibitem{Giddings:1989xe}
S.~B. Giddings, ``{Strings at the Hagedorn Temperature},''
  \href{http://dx.doi.org/10.1016/0370-2693(89)90288-8}{{\em Phys. Lett. B}
  {\bfseries 226} (1989) 55--61}.

\bibitem{Bowick:1989us}
M.~J. Bowick and S.~B. Giddings, ``{HIGH TEMPERATURE STRINGS},''
  \href{http://dx.doi.org/10.1016/0550-3213(89)90500-2}{{\em Nucl. Phys. B}
  {\bfseries 325} (1989) 631--646}.

\bibitem{Bagchi:2015nca}
A.~Bagchi, S.~Chakrabortty, and P.~Parekh, ``{Tensionless Strings from
  Worldsheet Symmetries},''
  \href{http://dx.doi.org/10.1007/JHEP01(2016)158}{{\em JHEP} {\bfseries 01}
  (2016) 158}, \href{http://arxiv.org/abs/1507.04361}{{\ttfamily
  arXiv:1507.04361 [hep-th]}}.

\bibitem{Garriga:1991tb}
J.~Garriga and A.~Vilenkin, ``{Quantum fluctuations on domain walls, strings
  and vacuum bubbles},'' \href{http://dx.doi.org/10.1103/PhysRevD.45.3469}{{\em
  Phys. Rev. D} {\bfseries 45} (1992) 3469--3486}.

\end{thebibliography}\endgroup


\end{document}